\documentclass[prl,reprint,superscriptaddress,%
amsmath,amssymb,aps,floatfix]{revtex4-2}

\usepackage{booktabs}
\usepackage{amsthm}
\usepackage[shortlabels]{enumitem}
\usepackage{braket}
\usepackage{comment}
\usepackage{graphicx, xcolor, varwidth}
\usepackage{color}
\usepackage{mathrsfs}
\usepackage{hyperref}
\DeclareMathOperator{\tr }{tr}

\newcommand{\kj}{\ket{\mathcal{J}}}
\newcommand{\bj}{\bra{\mathcal{J}}}

\renewcommand{\j}{\mathcal{J}}

\begin{document}
\title{Quantum State of a Gravitating Region}
\author{Raphael Bousso}
\affiliation{Leinweber Institute for Theoretical Physics and Department of Physics,\\
University of California, Berkeley, California 94720, U.S.A.}
\author{Sami Kaya}
\affiliation{Leinweber Institute for Theoretical Physics and Department of Physics,\\
University of California, Berkeley, California 94720, U.S.A.}
\author{Guanda Lin}
\affiliation{Leinweber Institute for Theoretical Physics and Department of Physics,\\
University of California, Berkeley, California 94720, U.S.A.}
\author{Arvin Shahbazi-Moghaddam}
\affiliation{Leinweber Institute for Theoretical Physics, Stanford, CA 94305, U.S.A.}
\begin{abstract}
We propose that any compact $d$-manifold with elliptic data, $\j$, prepares a quantum state $\kj$ on its finite $(d-1)$-boundary $\sigma$. Elliptic data consist of metric and field values, or their conjugates, but not both. No asymptotic structure is required. Inner products and traces are evaluated by the gravitational path integral with closed boundary conditions obtained by gluing elliptic data manifolds. In particular, we give a prescription for the R\'enyi entropies $S_n$ of a subregion $\chi$ of $\sigma$. In a class of examples, we find that $S_n$ is nonnegative and nonincreasing with $n$, as required for consistency. We obtain the von Neumann entropy by analytic continuation and find agreement with the minimal surface prescription of Bousso and Penington. 
\end{abstract}
\maketitle

\paragraph{Introduction}

The gravitational path integral (GPI) in $d+1$ dimensions can be used to approximate the partition function and the time evolution of a $d$-dimensional holographic boundary theory. Let $J$ denote a closed $d$-manifold together with elliptic boundary data. One evaluates the GPI by summing over $d+1$ manifolds $M$ with metric and fields that match $J$ on $\partial M$, the boundary of $M$. In the end, $J$ (more precisely, its embedding in the saddle geometries that dominate the GPI) is ``taken to infinity'' by scaling its data in a manner consistent with a targeted class of bulk manifolds, such as asymptotically flat~\cite{Gibbons:1976ue} or Anti-de Sitter (AdS)~\cite{Maldacena:1997re,Gubser:1998bc, Witten:1998qj} spacetimes.

\begin{figure}[t]
\centering
\includegraphics[width=\columnwidth]{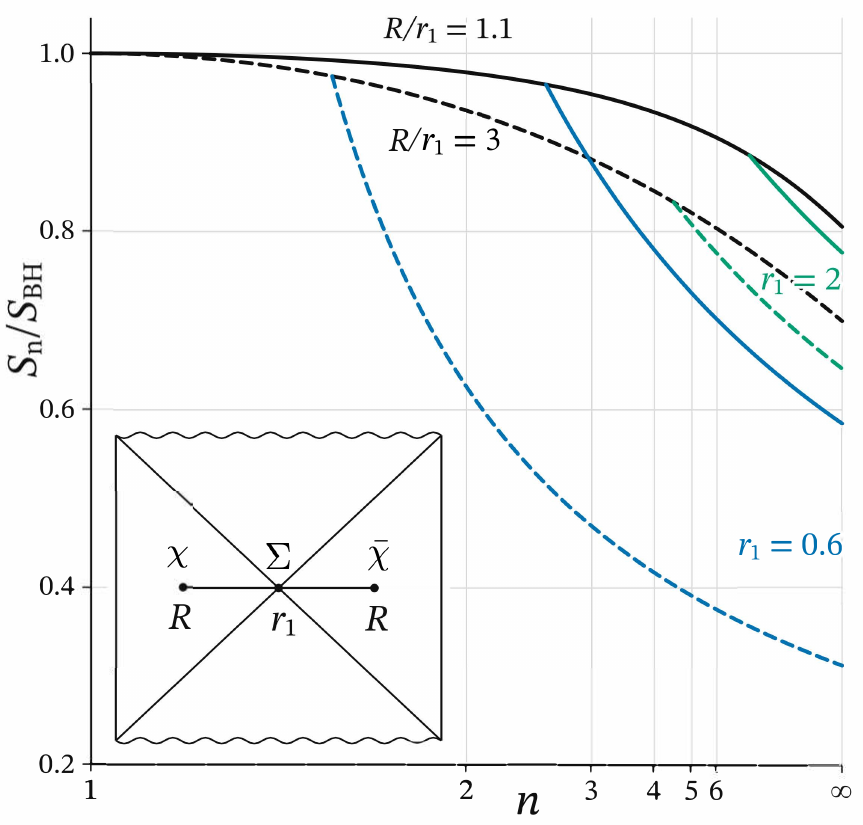}
\caption{R\'enyi entropies $S_n$ of the subregion $\chi$ of the boundary $\sigma=\chi\cup\bar\chi$ of an elliptic data manifold $\j$. $\j$ is obtained from the partial Cauchy slice $\Sigma$ (see Penrose diagram) by extracting mixed-conformal data; see Eq.~\ref{eq:mixed-modulus}. Cyclically gluing $2n$ copies of $\j$ yields a torus boundary condition for the GPI, from which $S_n$ is computed; see Fig.~\ref{fig:j-embeddings-btz-thermal}. We find that the von Neumann entropy $S_1=\lim_{n\to 1} S_n$ agrees with $S_{\rm BH}=2\pi r_1/(4G)$ for all $(r_1,R)$. This is consistent with the minimal surface prescription~\cite{Bousso:2022hlz}. -- Legend: for fixed $(r_1,R)$, at small $n$, $S_n/S_{\rm BH}$ depends only on the ratio $R/r_1$; the solid and dashed black curves show two examples. At large $n$, thermal AdS eventually dominates. This transition depends on $r_1$; examples are shown in blue and green. $S_n$ follows the lower (colored) branch.}
\label{fig:renyiplotmc}
\end{figure}

In this letter, we embrace a more general viewpoint that requires no asymptotic structure. Formally, the GPI makes equal sense whether or not the elliptic data are asymptotic. By removing this restriction, we aim to distill information about the structure of quantum gravity states in arbitrary spacetimes, including our universe. The same motivation underlies the recent proposal of generalized entanglement wedges~\cite{Bousso:2022hlz,Bousso:2023sya}, and we will find a concrete connection.

We propose that any compact $d$-manifold with elliptic data, $\j$, prepares an unnormalized pure quantum state $\kj$ on its $(d-1)$-dimensional \emph{boundary} $\sigma$. (Here we will only consider data $\j$ that fully specify a real Riemannian metric on $\sigma$.) The complex conjugate data $\j^\dagger$ prepares the corresponding bra $\bj$. A trace or inner product is computed by evaluating the GPI with a closed boundary condition $J$, formed by an appropriate gluing of elliptic data manifolds \footnote{As in all applications of the GPI, positivity properties are not guaranteed; thus the positivity and monotonicity of the R\'enyi entropies we compute furnish a nontrivial consistency check. We will discuss apparent violations of positivity~\cite{Maloney:2007ud,Keller:2014xba,Wall:2021bxi} in a companion paper.}.

An elliptic data manifold $\j$ may be obtained from a partial Cauchy slice $\Sigma$ of an arbitrary Lorentzian manifold, by stripping away half of the classical initial data on $\Sigma$. The map $\Sigma\!\to \!\j$ is not unique, and the quantum state $\kj$ need not be dominated by the classical Lorentzian geometry that develops from $\Sigma$. (With Dirichlet data, we will find explicit examples where it is not; this partly motivates our choice of ``mixed-conformal'' data in the latter portion of our analysis.)

\begin{figure*}[t]
\centering
\includegraphics[width=\textwidth]{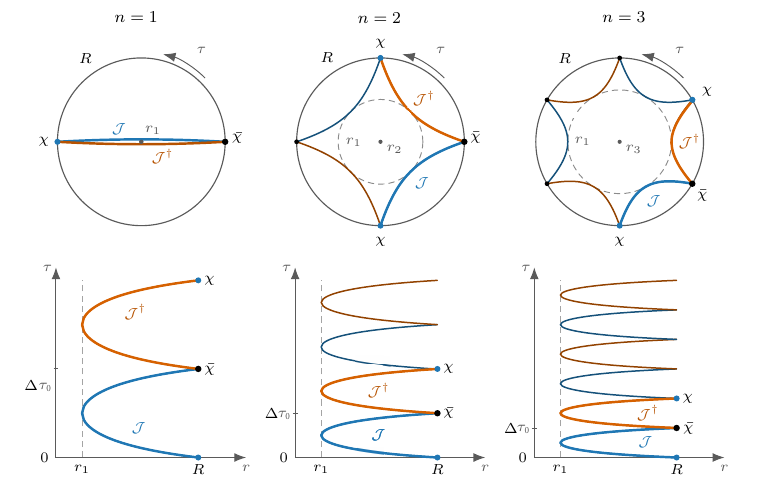}
\caption{Calculation of $\tr \tilde\rho_\chi^{\, n}$. We start with the partial Cauchy slice $\Sigma$ of a nonrotating BTZ black hole (Fig.~\ref{fig:renyiplotmc}), with $(r_1,R)= (2,2.8)$. In this example, the elliptic data manifold $\j$ retains the intrinsic metric -- the Dirichlet data -- of $\Sigma$, but not the extrinsic curvature. (Mixed-conformal data yield qualitatively similar embeddings but different R\'enyi entropies.) The partially glued pair $\j^\dagger \cup_{\bar\chi} \j$ represents the reduced density operator $\tilde \rho_\chi$. Cyclically gluing $n$ such pairs yields the closed elliptic data manifold $J_n$. We compute $\tr \tilde\rho_\chi^{\, n}$ by evaluating the GPI with boundary condition $J_n$. The graphs show the
embeddings of $J_n$ in the Euclidean BTZ (top) and thermal AdS (bottom) saddles. (In the thermal AdS saddle, identify $\tau\sim\tau+2n\Delta\tau_0$.)
The competition between saddles is reminiscent of the Hawking-Page transition~\cite{Hawking:1982dh,Witten:1998zw}, but $\j$ and $\chi$ are in no sense ``asymptotic.''}
\label{fig:j-embeddings-btz-thermal}
\end{figure*}

Our work generalizes and refines Refs.~\cite{Lewkowycz:2013nqa,Marolf:2020xie, Araujo-Regado:2022gvw, Chen:2025fwp}\footnote{Our perspective is distinct from each of these works. Our state preparation is not asymptotic~\cite{Lewkowycz:2013nqa,Marolf:2020xie} but finite. Our interest is not primarily in closed universes~\cite{Marolf:2020xie,Chen:2025fwp}. We do not presume any dual theory, such as the $T\bar T$-deformed CFT~\cite{McGough:2016lol, Hartman:2018tkw} that might be relevant in certain limiting settings~\cite{Araujo-Regado:2022gvw}. In particular, we compute exclusively using the GPI, never in a putative dual theory~\cite{Araujo-Regado:2022gvw} (see also~\cite{Soni:2024aop})}. A key novelty is our study of the entanglement structure of the state $\kj$ via the GPI, which in particular allows us to make contact with generalized entanglement wedges~\cite{Bousso:2022hlz}. For any finite-boundary subregion $\chi\subset \sigma$, we define a reduced state $\rho_\chi$, and we give a prescription for computing its R\'enyi entropies; see Fig.~\ref{fig:renyiplotmc}. (Implicitly the R\'enyi spectrum determines $\rho_\chi$.) We compute the von Neumann entropy of $\rho_\chi$ by analytic continuation, and we  will find that it agrees with~\cite{Bousso:2022hlz}.

\paragraph{R\'enyi entropy prescription}

Let $\j$ be an elliptic data manifold with boundary $\sigma$; see Fig.~\ref{fig:j-embeddings-btz-thermal}. The unnormalized density operator $\tilde \rho_\sigma = \kj\bj$ is represented by the pair $\j,\j^\dagger$. Given an open subregion $\chi\subset \sigma$ with complement $\bar\chi=\sigma\setminus\chi$, the partial trace of $\tilde\rho_\sigma$ over $\bar\chi$, $\tilde \rho_\chi = \tr_{\bar\chi}\tilde\rho_\sigma$, is performed by gluing together the pair $\j,\j^\dagger$ at $\bar\chi$. The result, denoted by $\j^\dagger\cup_{\bar\chi} \j$, is a data manifold whose boundary consists of two copies of $\chi$.
$\tilde \rho_\chi^{\, n}$ is prepared by gluing together $n$ copies of $\j^\dagger\cup_{\bar\chi} \j$ sequentially across $2n-2$ copies of $\chi$; and $\tr \tilde \rho_\chi^{\, n}$ is prepared by gluing together the remaining two copies of $\chi$. This yields a closed manifold $J_n$ that carries the alternating sequence of boundary conditions $\j^\dagger$, $\j$, $\ldots$, repeated $n$ times. In particular, the norm of $\kj$, $\tr \tilde \rho_\sigma$, corresponds to $J_1$ ($2$ copies of $\j$ glued together across all of $\sigma$).

For all positive integers $n$, $\tr \tilde \rho_\chi^{\, n}$ can be computed by evaluating the GPI with boundary $J_n$. In the saddle-point approximation,
\begin{equation}\label{eq:gpi}
    \tr \tilde\rho_\chi^{\, n} \approx \sum_s \exp\left(-I[g_{ij}^{(s)}]\right)\,,
\end{equation}
where $s$ labels classical solutions with $(d+1)$-dimensional metric $g_{ij}$ and boundary $J_n$, and $I$ is the Euclidean Einstein--Hilbert action with boundary terms appropriate to the specified data. Here we consider pure Einstein gravity for simplicity. With elliptic data consisting of the induced metric $h_{ab}$, the action is \cite{Gibbons:1976ue,York:1972sj,Hayward:1993my}
\begin{align}
I[g_{ij}]
& =-\frac{1}{16\pi G}
\int_{M} d^{d+1}x\,\sqrt{g}\,(R-2\Lambda)
\notag\\
&\hspace{2em}
-\frac{1}{8\pi G}
\int_{J_n} d^d x\,\sqrt{h}\,K
\notag\\
&\hspace{2em}
-\frac{1}{8\pi G}\sum_{\text{joints}}
\int d^{d-1}x\,\sqrt{\gamma}\,\Theta\,.
\label{eq-IEinstein_intro}
\end{align}
The sum in the Hayward term is over the $2n$ joint surfaces ($n$ copies each of $\bar\chi$ and $\chi$), at which the embedding of $J_n$ into $M$ can have a corner with exterior angle $\Theta$. With the mixed-conformal elliptic data defined below, the coefficient of the second (GHY) term is $-(8\pi d G)^{-1}$~\cite{York:1972sj}.

This prescription may be used to compute the $n$-th R\'enyi entropy of $\rho_\chi$,
\begin{equation}\label{eq:Renyiintro}
    S_n(\rho_\chi) = \frac{1}{1-n}\log \frac{\tr \tilde\rho_\chi^{\, n}}{(\tr \tilde\rho_\chi)^n}\,.
\end{equation}
The analytic continuation of the integer R\'enyi entropies $n>1$ to $n=1$ yields the von Neumann entropy of $\rho_\chi$.

\paragraph{Worked examples}
Let the partial Cauchy slice $\Sigma$ be the time-reflection-symmetric slice of a nonrotating BTZ black hole \cite{Banados:1992wn} of horizon radius $r_1$, cut off symmetrically at radius $R$ in the two exterior regions. Thus, $\sigma$ consists of two circles of length $2\pi R$. We choose $\chi$ to be one of those circles. We consider two different choices of elliptic data in the interior of $\Sigma$: first, the intrinsic metric $h_{ab}$ on $\Sigma$; and second, the conformal class $[h_{ab}]$ and the trace $K$ of the extrinsic curvature of $\Sigma$. 

This setting is chosen for simplicity. We stress that the asymptotically AdS structure of BTZ plays no role in our construction. We could equally have studied examples with nonnegative values of the cosmological constant. We shall use units in which the cosmological constant is $-1$. 

\paragraph{Dirichlet data}

Topologically, $\Sigma$ is a cylinder. We retain the intrinsic metric data $\j$ of $\Sigma$. It consists of two patches (the left and right exterior), each given by
\begin{equation}\label{eq:oneex}
    ds^2=\frac{dr^2}{r^2-r_1^2}+r^2d\phi^2\,,~~\phi\sim\phi+2\pi\,,~~ r_1\leq r\leq R
\end{equation}
and glued together at the horizon $r=r_1$. Thus, $J_n$ is the torus consisting of $2n$ copies of $\j$ glued cyclically across the boundary circles $\chi$ and $\bar\chi$.

The GPI with boundary condition $J_n$ has two (rotationally symmetric and replica-symmetric) saddle points. The first is a Euclidean BTZ solution,
\begin{equation}
ds^2=(r^2-r_n^2)d\tau^2+\frac{dr^2}{r^2-r_n^2}
+r^2d\phi^2\,,~~
\tau\sim\tau+\frac{2\pi}{r_n}\,.
\label{eq:dir-sym-btz}
\end{equation}
The embedding $\tau_n(r)$ of the metric in Eq.~\ref{eq:oneex} obeys
\begin{equation}\label{eq:dirbtzemb}
\frac{d\tau_n}{dr}
=
\frac{\sqrt{r_1^2-r_n^2}}
{(r^2-r_n^2)\sqrt{r^2-r_1^2}}\,.
\end{equation}
Replica closure around the Euclidean time circle gives
\begin{equation}
r_n=
\frac{r_1 \sin\!\left(\frac{\pi}{2n}\right)}
{\sqrt{1-\frac{r_1^2}{R^2}\cos^2\!\left(\frac{\pi}{2n}\right)}}\,,
\label{eq:dir-sym-rn}
\end{equation}
for $n>1$, and $r_n=r_1$ for $n=1$ by continuity.

The Einstein--Hilbert term and the Gibbons--Hawking--York term cancel in this solution. Thus only the Hayward terms contribute, at the two circles of radius $R$, where adjacent copies of $\j$ meet with exterior angle $2\theta_n$. Setting $r=R$ in Eq.~\ref{eq:dirbtzemb}, one finds
\begin{equation}
\cos\theta_n
=
\frac{\sqrt{r_1^2-r_n^2}}{\sqrt{R^2-r_n^2}}
=
\frac{r_1}{R}\cos\!\left(\frac{\pi}{2n}\right)\,.
\end{equation}
The $2n$ joints yield a total action
\begin{equation}
I_{n>1}^{\rm BTZ}
= -\frac{nR}{G}\,\theta_n
\,,~~~I_1^{\rm BTZ}=-\frac{\pi R}{2G}\,.
\end{equation}

The second saddle is vacuum AdS with Euclidean time period $\Delta\tau$,
\begin{equation}
\begin{split}
ds^2&=(1+r^2)d\tau^2+\frac{dr^2}{1+r^2}
+r^2d\phi^2\,,
\\
&\hspace{2em}
\phi\sim\phi+2\pi\,, \qquad \tau\sim\tau+\Delta\tau\,.
\end{split}
\label{eq:dir-sym-ads}
\end{equation}
The embedding $\tau_0(r)$ of the metric in Eq.~\ref{eq:oneex} satisfies
\begin{equation}\label{eq:diremb}
\frac{d\tau_0}{dr}
=
\frac{\sqrt{1+r_1^2}}
{(1+r^2)\sqrt{r^2-r_1^2}}\,,
\end{equation}
so one copy of $\j$ spans Euclidean time
\begin{equation}
\Delta\tau_0(R,r_1)
=
2\,\operatorname{arctanh}\!\left[
\frac{\sqrt{R^2-r_1^2}}{R\sqrt{1+r_1^2}}
\right]\,.
\end{equation}
Replica closure requires that $\Delta\tau= 2n\Delta\tau_0(R,r_1)$.

All terms in Eq.~\ref{eq-IEinstein_intro} contribute, yielding the total action
\begin{equation}
I_n^{\rm AdS}
= n I_1^{\rm AdS}\,;~~~
I_1^{\rm AdS}=
-\frac{\Delta\tau_0(R,r_1)}{2G}
-\frac{R}{G}\theta_0\,,
\label{eq:dir-sym-ads-action}
\end{equation}
where $\cos \theta_0=\sqrt{(1+r_1^2)/(1+R^2)}$ is the half exterior angle at each joint $r=R$. The BTZ saddle dominates if and only if
\begin{equation}
R(\theta_n-\theta_0)
> \frac{\Delta\tau_0(R,r_1)}{2}\,.
\label{eq:dir-sym-dominance}
\end{equation}
The angle $\theta_n$ decreases with $n$; in fact, for any $R,r_1$, the AdS saddle will dominate for sufficiently large $n$.

For $n>1$, the R\'enyi entropy is $S_n=(I_n^\star-nI_1^\star)/(n-1)$,
where $I_n^\star\equiv\min\{I_n^{\rm BTZ},I_n^{\rm AdS}\}. $ 
Depending on $(r_1,R)$, there are three regimes. 

For $r_1<0.6627...$, the thermal AdS saddle dominates already at $n=1$ if $R$ is sufficiently large. Then thermal AdS dominates for all larger $n$ as well, and
\begin{equation}\label{eq:s0}
    S_n=0~\text{for all}~n\geq 1\,.
\end{equation}
In this regime, the state $\kj$ does not correspond to the Lorentzian geometry from which the elliptic data manifold $\j$ was obtained \footnote{The thermal AdS-dominated regime at $n=1$ intersects with the $R\to\infty$ limit studied in Ref.~\cite{Araujo-Regado:2022gvw}, for small $r_1$, indicating that even \emph{global} Cauchy slice holography may not be compatible with using Dirichlet data.}. This motivates our choice of different elliptic data (mixed-conformal) below. A detailed study of the correspondence between states and finite Lorentzian domains will be presented elsewhere~\cite{FullPaper}.

If BTZ dominates at $n=1$ but thermal AdS dominates at a given $n>1$,
then
\begin{equation}
\begin{split}
S_n
&=
\frac{n}{G(n-1)}
\bigg[
R\left(\frac{\pi}{2}-\theta_0\right)
-
\frac{\Delta\tau_0(R,r_1)}{2}
\bigg]\,.
\end{split}
\label{eq:dir-sym-Renyi-ads}
\end{equation}

If BTZ dominates at a given $n>1$ (and hence at $n=1$), then
\begin{equation}
S_n
=
\frac{nR}{G(n-1)}
\left(\frac{\pi}{2}-\theta_n\right)\,;
\label{eq:dir-sym-Renyi-btz}
\end{equation}
and taking $n\to1$ gives the von Neumann entropy
\begin{equation}\label{eq:vNdir}
S_1=\frac{\pi r_1}{2G}=\frac{2\pi r_1}{4G}\,.
\end{equation}

We stress again that two phases are present at $n=1$. The von Neumann entropy 
changes discontinuously between $\pi r_1/(2G)$ and $0$.

\paragraph{Mixed-conformal data} We consider the same time-reflection-symmetric slice $\Sigma$. Instead of keeping its intrinsic metric $h_{ab}$, Eq.~\ref{eq:oneex}, we now retain the trace $K$ of the extrinsic curvature and the conformal class $[h_{ab}]$ in the interior of $\Sigma$. The conformal class of an annulus has only one modulus, $m$. Thus the elliptic data $\j$ are
\begin{equation}
m=\frac{1}{\pi r_1}
\arccos\frac{r_1}{R}\,;~~~K=0\,.
\label{eq:mixed-modulus}
\end{equation}
On the boundary $\sigma=\partial\j$, we retain the full $1$-metric (two circles $\chi,\bar\chi$ of length $2\pi R$); hence the qualifier ``mixed.'' The GPI boundary condition $J_n$ is the torus obtained by cyclically gluing $2n$ copies of $\j$ across $\chi,\bar\chi$.

Again there are two saddles for each $n$, with metrics in Eq.~\ref{eq:dir-sym-btz} and Eq.~\ref{eq:dir-sym-ads}. For the BTZ saddle, the embedding $\tilde\tau_n(r)$ of a $K=0$ face satisfies
\begin{equation}
\frac{d\tilde\tau_n}{dr}=
\frac{E_n}{(r^2-r_n^2)\sqrt{r^4-r_n^2r^2-E_n^2}}\,,
\end{equation}
where $E_n>0$ for integer $n>1$. The turning circle obeys $r_{\min,n}^2
=
\left(
r_n^2+\sqrt{r_n^4+4E_n^2}
\right)/2$. The modulus condition and replica closure imply
\begin{align}
m
&=
\frac{1}{\pi}
\int_{r_{\min,n}}^R
\frac{dr}{\sqrt{r^4-r_n^2r^2-E_n^2}}\,,
\label{eq:mixed-btz-modulus}
\\
\frac{\pi}{n r_n}
&=
2\int_{r_{\min,n}}^R
\frac{E_n\,dr}
{(r^2-r_n^2)\sqrt{r^4-r_n^2r^2-E_n^2}}\,.
\label{eq:mixed-btz-closure}
\end{align}
Together these determine $(r_n,E_n)$, though we will not provide closed-form expressions. At $n=1$, the solution is understood as the
folded limit $r_n\to r_1$, $E_n\to0^+$; the second equation has a singular
contribution from the neighborhood of the turning circle in this limit.

The GHY term vanishes since $K=0$. The action is
\begin{equation}
I_n^{\rm BTZ}
=
\frac{n}{G}
\left[
\pi m E_n-R\vartheta_n
\right]\,,~~~
I_1^{\rm BTZ}=-\frac{\pi R}{2G}\,,
\label{eq:mixed-btz-action}
\end{equation}
where the exterior angle $2\vartheta_n$ at each joint obeys
\begin{equation}
\cos \vartheta_n
=
\frac{E_n}{R\sqrt{R^2-r_n^2}}\,,
\qquad 0\le\vartheta_n\le\frac{\pi}{2}\,.
\label{eq:mixed-btz-angle}
\end{equation}

The embedding $\tilde\tau_0(r)$ of a $K=0$ face in the thermal AdS metric in Eq.~\ref{eq:dir-sym-ads} satisfies
\begin{equation}
\frac{d\tilde\tau_0}{dr}
=
\frac{E_0}
{(1+r^2)\sqrt{r^4+r^2-E_0^2}}\,,
\end{equation}
with turning radius
$r_{\min,0}^2=
\left(-1+\sqrt{1+4E_0^2}\right)/2$.
The modulus condition fixes $E_0$:
\begin{equation}
m
=
\frac{1}{\pi}
\int_{r_{\min,0}}^R
\frac{dr}{\sqrt{r^4+r^2-E_0^2}}\,.
\label{eq:mixed-ads-modulus}
\end{equation}
One copy spans Euclidean time
\begin{equation}
\Delta\tilde\tau_0
=
2\int_{r_{\min,0}}^R
\frac{E_0\,dr}
{(1+r^2)\sqrt{r^4+r^2-E_0^2}}\,,
\end{equation}
so replica closure sets the thermal period in Eq.~\ref{eq:dir-sym-ads} to $\Delta\tau= 2n\Delta\tilde\tau_0$. The bulk action is
$n[E_0(2\pi m)-\Delta\tilde\tau_0]/(2G)$. Adding the Hayward terms, we obtain
\begin{equation}
I_n^{\rm AdS}
= n I_1^{\rm AdS}\,;~
I_1^{\rm AdS}=
\frac{1}{G}
\left[
\pi m E_0-\frac{\Delta\tilde\tau_0}{2}-R\vartheta_0
\right]\,,
\label{eq:mixed-ads-action}
\end{equation}
where the exterior angle $2\vartheta_0$ at each joint satisfies
\begin{equation}
\cos\vartheta_0
=
\frac{E_0}{R\sqrt{1+R^2}}\,,
\qquad 0\le\vartheta_0\le\frac{\pi}{2}\,.
\end{equation}

The BTZ saddle dominates if and only if
\begin{equation}
\pi m(E_n-E_0)+\frac{\Delta\tilde\tau_0}{2}
<
R(\vartheta_n-\vartheta_0)\,.
\label{eq:mixed-dominance}
\end{equation}
For any physical value of the parameters, $0<r_1<R$, this condition is satisfied near $n=1$ and fails as $n\to\infty$, with the difference behaving monotonically in between. If thermal AdS dominates at a given $n>1$, then
\begin{equation}
S_n
=
\frac{n}{G(n-1)}
\left[
\pi m E_0-\frac{\Delta\tilde\tau_0}{2}+R\left(\frac{\pi}{2}-\vartheta_0\right)
\right]\,;
\label{eq:mixed-Renyi-ads}
\end{equation}
if BTZ dominates, then
\begin{equation}
S_n
=
\frac{n}{G(n-1)}
\left[
\pi m E_n+R\left(\frac{\pi}{2}-\vartheta_n\right)
\right]\,.
\label{eq:mixed-Renyi-btz}
\end{equation}
In the $n\to 1$ limit, Eq.~\ref{eq:mixed-Renyi-btz} applies regardless of $(r_1, R)$, and we obtain the $R$-independent von Neumann entropy
\begin{equation}\label{eq:mcvn}
S_1=\frac{\pi r_1}{2G}=\frac{2\pi r_1}{4G}\,.
\end{equation}

\paragraph{Discussion}

We computed the R\'enyi entropies of a boundary subregion $\chi\subset\sigma$  for two families of elliptic data manifolds $\j$ parameterized by $(r_1,R)$.
Both families were obtained from the time-reflection-symmetric partial Cauchy slice $\Sigma$ of a BTZ black hole of radius $r_1$ cropped at radius $R$. For the ``Dirichlet'' family, we retained the intrinsic metric of $\Sigma$; for the ``mixed-conformal'' family, we kept only the conformal class of the metric but added the trace of the extrinsic curvature of $\Sigma$ to the data.

For fixed $(r_1,R)$, in both families, we found that the R\'enyi entropies $S_n$ are nonnegative and nonincreasing in the range $1< n<\infty$. (This can be verified analytically even for mixed-conformal boundary conditions, where the $S_n$ are not available in closed form.) Thus our proposal passes an important nontrivial consistency check; see Fig.~\ref{fig:renyiplotmc}. 

For mixed-conformal data, the original BTZ geometry dominates the path integral near $n=1$ for all $(r_1,R)$. The von Neumann entropy $S_1(\chi)$, Eq.~\ref{eq:mcvn}, agrees with the Bekenstein-Hawking entropy of the minimal surface homologous to $\chi$ on $\Sigma$. This matches the conjecture~\cite{Bousso:2022hlz} that the state $\rho_\chi$ on $\chi$ should be dual to an entanglement wedge constructed by a minimal-area prescription. We regard this as a highly nontrivial result.

For Dirichlet data, there exists a regime in the $(r_1,R)$ plane where the dominant saddle near $n=1$ is \emph{not} the original BTZ solution from which $\j$ was extracted. This was expected but highlights an important conceptual question. If the same $\j$ can be extracted from a different Lorentzian spacetime -- or, as in our examples of thermal AdS dominance at $n=1$, from a different Euclidean solution -- then how should we construct a quantum state that uniquely captures a particular Lorentzian domain? 

We confront this issue in a companion paper~\cite{FullPaper} that considers also time-dependent examples. When \emph{mixed-conformal} data are extracted from a small \emph{complex deformation} of the initial data on $\Sigma$, the original geometry dominates in all examples. These results align with broad evidence favoring conformal data over Dirichlet data in GPI applications~\cite{Witten:2018lgb}. Moreover, we find that the von Neumann entropy agrees with the maximin prescription~\cite{Wall:2012uf,Bousso:2023sya}, not with the minimal area on $\Sigma$. Thus, our proposal, while reminiscent of tensor network models of holography in connecting geometries to states, does not suffer from their limitations in time-dependent settings.

Looking farther ahead, our proposal might furnish a derivation of the generalized quantum extremal surface prescription~\cite{Bousso:2022hlz,Bousso:2023sya,Bousso:2024iry,Bousso:2025fgg} from the GPI, analogous to the proof~\cite{Lewkowycz:2013nqa} of its asymptotically-AdS version~\cite{Ryu:2006bv,Hubeny:2007xt,Engelhardt:2014gca}\footnote{In general, the elliptic data manifold $\j$ for the R\'enyi calculation should be extracted from a Cauchy slice $\Sigma$ of $a'$, the spacelike complement of the ``input region'' $a$ of Ref.~\cite{Bousso:2023sya}. The input region itself does not appear in the calculation. This viewpoint is distinct from Refs.~\cite{Balasubramanian:2023dpj, Kaya:2025vof}.}. 
The construction of \cite{Lewkowycz:2013nqa} requires a geometric representation of the CFT density operator as an asymptotic boundary condition. The finite elliptic data manifold $\j^\dagger\cup_{\bar\chi}\j$ provides a natural analogue for a proof of~\cite{Bousso:2022hlz,Bousso:2023sya}. The extension of Ref.~\cite{Lewkowycz:2013nqa} to generalized entanglement wedges is not straightforward, however, because cases arise that have no parallel in the asymptotic setting.

\paragraph{Acknowledgments} We thank Gabriele Di Ubaldo, Xi Dong, Luca Iliesiu, Don Marolf, Henry Maxfield, Geoff Penington, Pratik Rath, Misha Usatyuk, and Aron Wall for insightful discussions. This work was supported in part by the Leinweber Institutes for Theoretical Physics at UC Berkeley and Stanford; and by the Department of Energy, Office of Science, Office of High Energy Physics through award DE-SC0025293 and QuantISED award DE-SC0019380. ASM is supported by the Department of Energy through awards DE-SC0019380 and DE-FOA0002563, by AFOSR award FA9550-22-1-0098, and by a Sloan Fellowship.

\bibliography{Refs}

\end{document}